\documentclass[aps,prb,twocolumn,showpacs,superscriptaddress,amsmath,amssymb,longbibliography]{revtex4-1}
\usepackage{graphicx}
\usepackage{dcolumn}
\usepackage{bm}
\usepackage[hidelinks,colorlinks=true,linkcolor=blue,citecolor=blue]{hyperref}
\usepackage[mathlines]{lineno}

\usepackage{braket}
\usepackage{color}
\usepackage{float}
\usepackage{amsmath}

\begin{document}
\title{Emergent Continuous Time Crystal in Dissipative Quantum Spin System without Driving}

\affiliation{Graduate School of China Academy of Engineering Physics, Beijing 100193, China}
\affiliation{Shenzhen Key Laboratory of Ultraintense Laser and Advanced Material Technology, Center for Intense Laser Application Technology, and College of Engineering Physics, Shenzhen Technology University, Shenzhen 518118, China}
\affiliation{Quantum Science Center of Guangdong-Hongkong-Macao Greater Bay Area (Guangdong), Shenzhen 518045, China}
\affiliation{Department of Physics, Renmin University of China, Beijing 100872, China}

\author{Shu Yang}
\affiliation{Graduate School of China Academy of Engineering Physics, Beijing 100193, China}
\affiliation{Shenzhen Key Laboratory of Ultraintense Laser and Advanced Material Technology, Center for Intense Laser Application Technology, and College of Engineering Physics, Shenzhen Technology University, Shenzhen 518118, China}

\author{Zeqing Wang}
\affiliation{Quantum Science Center of Guangdong-Hongkong-Macao Greater Bay Area (Guangdong), Shenzhen 518045, China}
\affiliation{Shenzhen Key Laboratory of Ultraintense Laser and Advanced Material Technology, Center for Intense Laser Application Technology, and College of Engineering Physics, Shenzhen Technology University, Shenzhen 518118, China}
\affiliation{Department of Physics, Renmin University of China, Beijing 100872, China}

\author{Libin Fu} 
\email{Corresponding author: lbfu@gscaep.ac.cn}
\affiliation{Graduate School of China Academy of Engineering Physics, Beijing 100193, China}
  
\author{Jianwen Jie}
 \email{Corresponding author: Jianwen.Jie1990@gmail.com}
\affiliation{Shenzhen Key Laboratory of Ultraintense Laser and Advanced Material Technology, Center for Intense Laser Application Technology, and College of Engineering Physics, Shenzhen Technology University, Shenzhen 518118, China}

\date{\today}
\begin{abstract}
Time crystals are a nonequilibrium phase of matter that extend fundamental spontaneous symmetry breaking into the temporal dimension, typically requiring external driving for their realization. Here, we explore the nonequilibrium phase diagram of a two-dimensional dissipative Heisenberg spin system without external coherent or incoherent driving. Through numerical analysis of spin dynamics, we identify nonstationary steady states, some of which are limit cycles with persistent periodic oscillations, while others exhibit chaotic, aperiodic behavior. The emergence of limit cycle steady states breaks the continuous time-translation symmetry of this time-independent many-body system, classifying them as continuous time crystals. We further validate these oscillatory behaviors by testing their stability against local perturbations and assess the robustness of the emergent continuous time crystals by introducing isotropic Gaussian white noise. This work provides insights into the intricate interplay between the dissipation and spin interaction, and opens possibilities for realizing dissipation-induced, heating-immune time crystals.
\end{abstract}

\maketitle
\section*{Introduction}
A time crystal (TC) is a phase of matter in which time-translation symmetry is spontaneously broken, and this spontaneous breaking manifests as persistent periodic oscillations in local physical quantities that are not inherent to the system's properties. This concept was originally proposed by Frank Wilczek as a phenomenon that spontaneously breaks time-translation symmetry  in the ground or equilibrium state \cite{PRL2012CTC,PRL2012QTC,PRL2012Ion,Physics2012}. Although this idea was later negated by a series of rigorous no-go theorems \cite{PRL2013Bruno,Nozieres_2013,PRL2015Wat} , which ruled out the possibility of TCs in closed quantum systems with only short-range interactions \cite{PRL2019longrange}, the introduction of time-dependent periodic coherent driving has enabled the realization of discrete time-translation symmetry breaking  in nonequilibrium setups \cite{PRA2015DTTS,PRL2016Khemani,PRB2016von}. This phenomenon, known as Floquet time crystal or discrete time crystal \cite{PRL2016FTC_1,PRL2017FTC_1,Physics2017,PRB2017FTCLMG,PRB2019FTCCM,PRL2021QSN,PRL2018NMR_theory,Review2020DTC,PRB2023Xu,chen2023robustlargeperioddiscretetime}, is characterized by a subharmonic response to the driving frequency and has been experimentally demonstrated in various systems \cite{PRL2018NMR,PRB2018NMR,choi2017observation}, such as trapped-ions \cite{zhang2017observation,Science2021ion} and superconducting qubits \cite{mi2022time}. Its key challenge remains the heating-induced limited lifetime, despite the proposal of mechanisms \cite{PRX2017PTC,Science2021ion,PRB2018PTC,PRL2021PTC,PRL2021PTC_2} such as many-body localization \cite{Science2021MBLDTC,mi2022time}, prethermalization \cite{PRX2017PTC,Science2021ion,PRB2018PTC,PRL2021PTC,PRL2021PTC_2}, and dissipation \cite{PRX2017PTC,Review2020DTC},  which aim to slow down thermalization.

Dissipation has been shown to serve as a resource for performing quantum tasks \cite{verstraete2009quantum,PRL1996Engineering,preskill2022,lin2013dissipative,shankar2013autonomously,PRL2018QRe,PRL2022OP,PRL2024Liu}. For instance, by {carefully} designing dissipation to compete with coherent driving \cite{Physics2021,PRL2018Gong,Zhu_2019,PRR2020DissFTC,RieraCampeny2020timecrystallinityin,PRB2021DissFTC,PRL2019Jaksch,NC2019Jaksch}, or by engineering competition among dissipative processes, including incoherent driving \cite{minganti2020correspondence}, one can achieve dissipative {time crystals}. {By transforming to a suitable rotating frame where the time-dependent aspect of the driving is eliminated, the system can exhibit a continuous time crystal (CTC) characterized by the spontaneous breaking of continuous time-translation symmetry \cite{PRL2022seeding,PRL2023CTC,PRA2019dissTC,Tucker_2018,PRB2019DissCTC,PRA2020DissCTC,PRB2021dbtc,PRL2018BTC}.} These dissipative {time crystals} have been experimentally observed in various systems \cite{PRL2021dissDTC,PRL2021Kongkhambut,Science2022CTC,Science2019Nishant,PRL2019Zupancic,Dreon2022Nature,taheri2022all,PRL2019CTCBEC,PRB2019CTC}. {A perfect time crystal} should exhibit {an} infinite lifetime 
\cite{Sacha_2018,khemani2019brief,RMP2023Norman,Guo_2020}, {characterized by} a robust never-ending oscillatory (OSC) {behavior.} This mathematically corresponds to a stable closed trajectory in phase space, defined as {a} limit cycle (LC), a core element in nonlinear {dynamical phenomena} \cite{strogatz2018nonlinear}, such as {in grid power dynamics} \cite{KASIS2021109736}, circadian clocks \cite{Gonze2011712729} and quantum synchronization \cite{PRL2017kerr,PRA2018QSB,PRL2018,PRA2019tribit,PRL2020exp,PRA2020two,PRR2023Zhang,wang2023absence}. 
 LCs are a broader concept than TCs; they do not require the system to be either a single-body or a many-body system, nor do they necessarily imply the breaking of time-translation symmetry. The quest for nonequilibrium states and phases \cite{PRA2023Nie,breuer2002theory,rivas2011open,RMP2017open,RMP2021open,CMF2016PRX,Phase2013PRL_Lee,Jin2013PRL,SP2013PRL,PRX2017SPE,Jin2023PRB,
PRL2022Zhang,PRL2022Li,PRL2023RB,PRA2023Jin,PRA2023Haga,Lee2011PRA,PRA2016Wilson,PRL2016Schir,PRA2018driven,
Parmee_2020,PRA2012Qian,Owen2018,PRA2015driven,PRA2023Lixing,
PRB2022pass} that include OSC phases \cite{PRA2015driven,PRA2023Lixing,PRB2022pass,Lee2011PRA,
PRA2015driven,PRA2016Wilson,PRL2016Schir,PRA2018driven,Owen2018,Parmee_2020,PRA2012Qian,bunkov2012spinsuperfluiditymagnonbec} is a fundamentally important task in physics, predating TCs.

To the best of our knowledge, {all systems} exhibiting {time crystal} behavior are subjected to external driving, whether it be Floquet-driven \cite{PRA2015DTTS,PRL2016Khemani,PRB2016von,PRL2016FTC_1,PRL2017FTC_1,
Physics2017,PRB2017FTCLMG,PRB2019FTCCM,PRL2021QSN,PRL2018NMR_theory,Review2020DTC}, incoherent-driven \cite{minganti2020correspondence}, or driven-dissipative \cite{PRL2021Kongkhambut,Physics2021,PRL2018Gong,PRL2018BTC,PRL2018Gong,Zhu_2019,PRR2020DissFTC,
RieraCampeny2020timecrystallinityin,PRB2021DissFTC,PRL2022seeding,PRL2023CTC,PRA2019dissTC,
Tucker_2018,PRB2019DissCTC,PRA2020DissCTC,PRB2021dbtc,PRL2021dissDTC,Science2022CTC,
taheri2022all,PRL2019CTCBEC,PRB2019CTC,Science2019Nishant,PRL2019Zupancic,Dreon2022Nature} scenarios. However, external driving not only {introduces} complexity to the real system but also raises concerns about heating. Moreover, the presence of driving {complicates the discernment of} the contributions of other factors, such as interaction and dissipation, to the OSC behavior. Therefore, a fundamental question arises: Can a purely dissipative quantum system, {without external coherent or incoherent driving}, exhibit OSC behaviors? If so, {could} these oscillations be {time crystals}? {Positive} answers to these questions {could pave the way for} realizing dissipation-induced, {heating-immune time crystals}.

We address this question by theoretically {unveiling} the nonequilibrium steady-state phase diagram of a dissipative Heisenberg spin system, specifically one without external driving. We {demonstrate} the {emergence} of LC and chaotic steady-states, supported by spin dynamics and linear stability analysis. We {characterize the LC steady states} as a CTC by examining {their} robustness to noisy interactions and dissipation in the thermodynamic limit.

\begin{figure}[t]
\centering
\includegraphics[width=1.0\linewidth]{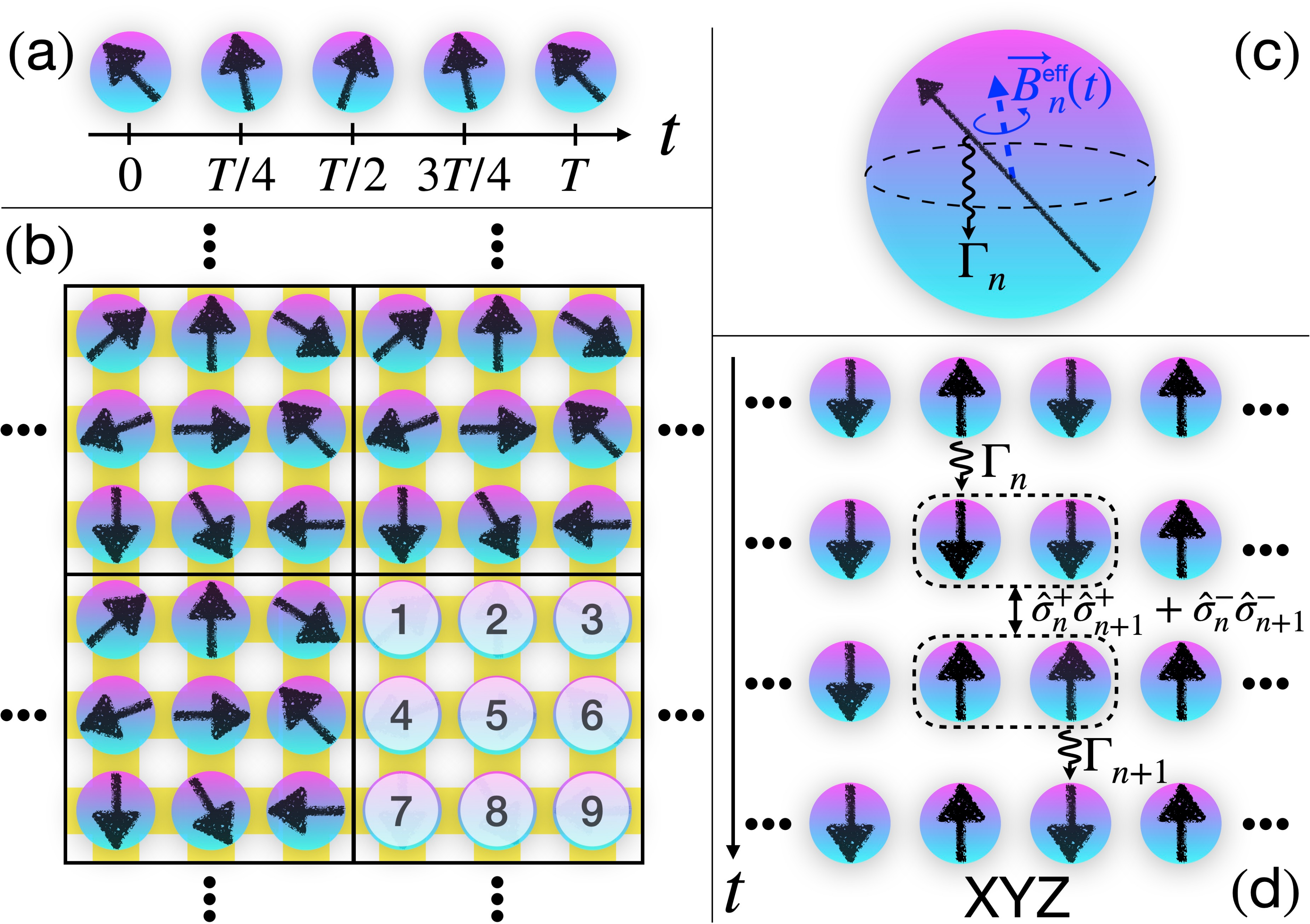}
\caption{\label{fig1}{\bf Illustration of continuous time crystals (CTCs) and their mechanism.} A persistent oscillation with period \( T \) of spins in \textbf{a} is the core feature of the CTCs, which emerge in a two-dimensional dissipative Heisenberg spin system composed of infinite \( 3 \times 3 \) clusters, with each spin indexed by \( n \in [1, 9] \) in \textbf{b}. As shown in \textbf{c}, the dynamics of the \( n \)th spin are governed by the interplay between the dissipation \( \Gamma_n \), which drives the spin downward, and the  XYZ  spin interaction, which induces an effective magnetic field \( {\vec{B}}^{\text{eff}}_n \). \textbf{d} illustrates the mechanism of the CTCs, realized through a microscopic cycle that involves three steps: the energy loss induced by dissipation \( \Gamma_n \), spin fluctuations driven by the anisotropic interaction \( \hat{\sigma}^{+}_n \hat{\sigma}^{+}_{n+1} + \hat{\sigma}^{-}_n \hat{\sigma}^{-}_{n+1} \), and the energy gain induced by dissipation \( \Gamma_{n+1} \).}
\end{figure}

\section*{Results}
\subsection*{System}
{We consider a dissipative interacting system governed by the Lindblad master equation  ($\hbar=1$), 
\begin{align}
\label{lindblad}
 \frac{d\hat{\rho}(t)}{dt}= -i\left[\hat{H}, \hat{\rho}(t)\right] +\frac{1}{2}\sum_n \mathcal{D}_{n}[\hat \rho(t)],
\end{align}
where $\hat{H}$ is the Hamiltonian governing {the} system’s unitary evolution. 

We consider {the} Hamiltonian {to be the } paradigmatic spin-1/2 Heisenberg XYZ model,
\begin{eqnarray}\label{Hamiltonian}
	\hat{H}=\sum_{\langle mn\rangle}\hat{V}_{mn}=\sum_{\langle mn\rangle}\frac{1}{2d}\sum_{\alpha=x,y,z}J_\alpha\hat{\sigma}^\alpha_m\hat{\sigma}^\alpha_n,
\end{eqnarray}
where $\hat{\sigma}_{n}^{\alpha}$ is the Pauli operator for $n$th spin and all spins are localized in a $d$-dimensional cubic lattice. The nearest-neighbor spin pairs, {denoted} by $\langle mn\rangle$, are 
anisotropic interacting with strength $J_\alpha$. Each spin undergoes an incoherent {downward flipping process} governed by  
\begin{eqnarray}\label{Dj}
\mathcal{D}_{n}[\hat{\rho}] = \Gamma\left(\hat{\sigma}^{-}_{n}\hat{\rho} \hat{\sigma}^{+}_{n} - \{\hat{\sigma}^{+}_{n}\hat{\sigma}^{-}_{n} , \hat{\rho} \}/2\right),
\end{eqnarray}
at strength $\Gamma$, with $\hat{\sigma}^{\pm}_{n}=(\hat{\sigma}^{x}_{n}\pm i\hat{\sigma}^{y}_{n})/2$. 

These single-site dissipations break the time reversibility and other relevant symmetries of the system which usually results in the uniqueness of a stationary steady state \cite{evans1977irreducible,Frigerio_1978,Prosen_2012,Prosen_2015,PRA2010Schirmer}.
{As illustrated in Fig. \ref{fig1}(a), a CTC is a nonstationary steady state characterized by a periodic oscillation in the local observable $\hat{O}$ of a time-independent many-body system, which spontaneously breaks the continuous time-translation symmetry of the system. This periodic oscillation is stabilized by many-body interactions in the thermodynamic limit, denoted by $\lim_{N \rightarrow \infty} \langle \hat{O} \hat{\rho}_{ss}(t \gg 1) \rangle = O(t) = O(t+T)$, where $N$ is the system size and $T$ is the period of oscillation. Moreover, the periodic oscillation is robust against small, locality-preserving perturbations to both the state and the equation of motion \cite{PRB2017TCdefinition, khemani2019brief, RMP2023Norman}.} The above definition of CTC} excludes the LC behaviors exhibited by the nonlinear systems with only a few degrees of freedom, such as a single van der Pol oscillator \cite{pikovsky2001synchronization} and Belousov–Zhabotinsky reaction \cite{enns2012nonlinear}, from being considered as candidates for time crystals.
 
\begin{figure*}
\centering
\includegraphics[width=17cm]{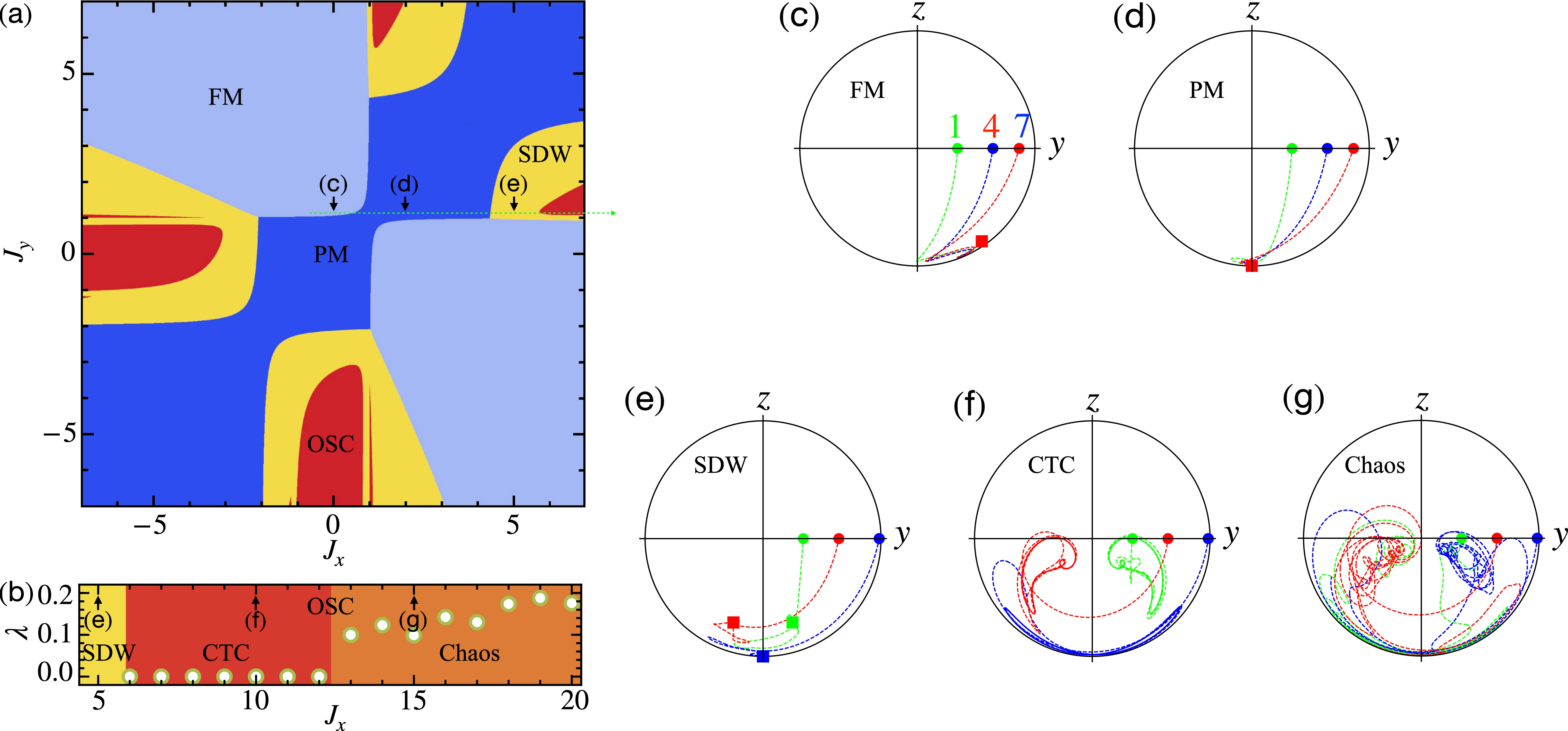}
\caption{\label{fig2}{\bf Nonequilibrium phase diagram and spin dynamics.} \textbf{a} shows the mean-field phase diagram with \( J_z = \gamma = 1 \), which includes three stationary phases: paramagnetic (PM), ferromagnetic (FM), and spin-density-wave (SDW), along with one additional nonstationary oscillatory (OSC) phase. The OSC phases can further be distinguished into continuous time crystal (CTC) and chaotic phases by calculating the Lyapunov exponent \( \lambda \), as exemplified in \textbf{b} with \( J_y = 1.1 \) (marked with the green dotted line in {\bf a}). \textbf{c-g}, indicated by arrows in \textbf{a-b} and corresponding to \( J_x = 0, 2, 5, 10, \) and \( 15 \), respectively, display the magnetization dynamic trajectories on the \( y \)-\( z \) plane for spins with \( n = 1, 4, 7 \) in each phase. The solid circles mark the initial states and the fixed-point fates of FM, PM, and SDW are marked with square blocks.}
\end{figure*}

\subsection*{Mean-field phase diagram}
To {investigate} the rich nonequilibrium phases in the thermodynamic limit ($N\rightarrow\infty$) \cite{Phase2013PRL_Lee}, we conduct our study at {the} mean-field level which assumes that all spins are uncorrelated, meaning the system is in a product state: $\hat{\rho}\approx\otimes_{n=1}^{N\rightarrow\infty}\hat{\rho}_{n}$. 
{Given} the limitations of the mean field theory in one-dimensional systems \cite{Phase2013PRL_Lee,CMF2016PRX}, we focus on two-dimensional systems {composed of} infinite $3\times3$ clusters as illustrated in Fig. \ref{fig1}(b).  {It is important to note that the cluster here refers to a unit cell in which the spins can exist in distinct states. The system is assumed to consist of an infinite number of such clusters, and the boundaries of these clusters are treated using periodic boundary conditions.}  In other words, we rewrite {the} density matrix as $\hat{\rho}\approx\hat{\rho}_{C}\otimes\hat{\rho}_{C}\otimes\hat{\rho}_{C}\cdots$ with $\hat{\rho}_{C}=\otimes_{n=1}^{N_c}\hat{\rho}_{n}$ and a cluster size $N_{C}=9$. Therefore, a set of $3N_{c}$ nonlinear Bloch equations is obtained from {Eq. (\ref{lindblad})},

\begin{align}\label{BEs}
\frac{d\langle\hat{\sigma}^x_n\rangle}{dt}&=-\frac{\Gamma}{2}\langle\hat{\sigma}^x_n\rangle+\sum_{m}\frac{J_y\langle\hat{\sigma}^z_n\rangle\langle\hat{\sigma}^y_m\rangle-J_z\langle\hat{\sigma}^y_n\rangle\langle\hat{\sigma}^z_m\rangle}{d},\nonumber\\
\frac{d\langle\hat{\sigma}^y_n\rangle}{dt}&=-\frac{\Gamma}{2}\langle\hat{\sigma}^y_n\rangle+\sum_m\frac{J_z\langle\hat{\sigma}^x_n\rangle\langle\hat{\sigma}^z_m\rangle-J_x\langle\hat{\sigma}^z_n\rangle\langle\hat{\sigma}^x_m\rangle}{d},  \\
\frac{d\langle\hat{\sigma}^z_n\rangle}{dt}&=-\Gamma(\langle\hat{\sigma}^z_n\rangle+1)+\sum_{m}\frac{J_x\langle\hat{\sigma}^y_n\rangle\langle\hat{\sigma}^x_m\rangle-J_y\langle\hat{\sigma}^x_n\rangle\langle\hat{\sigma}^y_m\rangle}{d},\nonumber
\end{align}
where $n$ is an index within a cluster $C$ and the sum over $m$ encompasses the nearest neighbors of the $n$th spin. Although the mean-field approach may fail in systems with short-range interactions, for the nearest-neighbor Heisenberg spin system considered in our study, we conducted precise full quantum numerical calculations on finite-size systems with periodic boundary conditions. These calculations reveal that two-body correlations significantly diminish as the system size increases (see Supplementary Note 1). This observation is consistent with the approximations inherent in mean-field methods.

Figure \ref{fig2}(a) shows the phase diagram numerically determined from spin dynamics {with fixed $J_{z}=\gamma=1$}. The paramagnetic {(PM)} phase aligns all spins {downward, i.e.,} $\langle\hat{\sigma}_{n}^{x(y)}\rangle=0,\langle\hat{\sigma}_{n}^{z}\rangle=-1$, preserving the system's $Z_2$ symmetry ($\hat{\sigma}_{n}^{x}\rightarrow-\hat{\sigma}_{n}^{x}$, $\hat{\sigma}_{n}^{y}\rightarrow-\hat{\sigma}_{n}^{y}$).  This $Z_2$ symmetry of the system is spontaneously broken in the ferromagnetic (FM) phase, where $\langle\hat{\sigma}_{n}^{x(y)}\rangle\neq0$. The spin-density-wave (SDW) phase has a period greater than two lattice sites in at least one direction. The antiferromagnetic (AFM) and staggering XY (sXY) phases as found in Lee \textit{et al.} \cite{Phase2013PRL_Lee} would emerge when the selected cluster contains an even number of spins along any axis (see Supplementary Note 2).  In addition to those stationary phases, a nonstationary OSC phase, which spontaneously breaks time-translation symmetry, emerges within SDW phase region. Unlike PM and FM, which are spatially uniform phases and can exist without dissipation \cite{Phase2013PRL_Lee,sachdev_2011}, SDW and OSC are spatially non-uniform states that exist only under nonequilibrium circumstances. 

\begin{figure*}
\centering
\includegraphics[width=17.0cm]{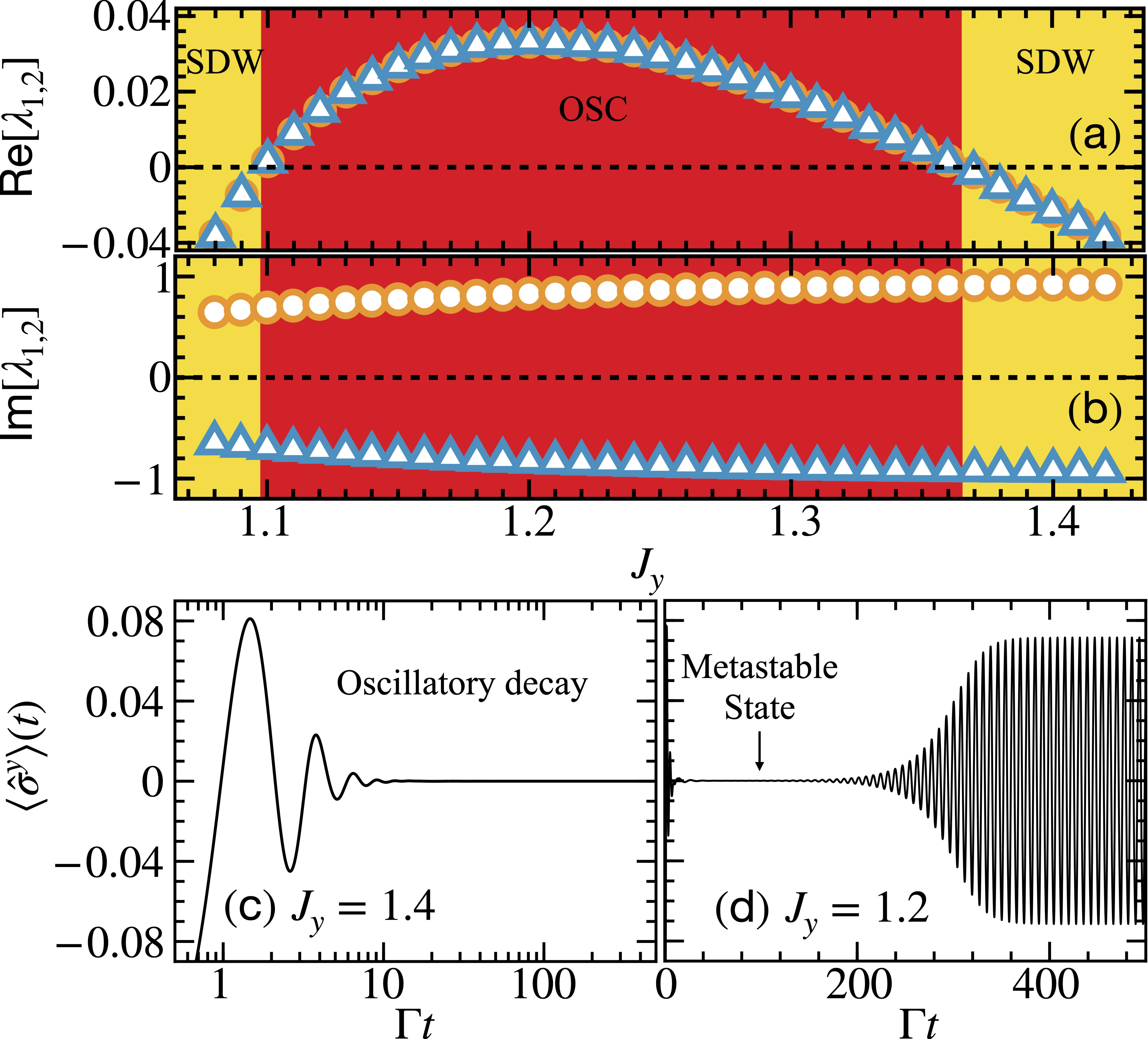}
\caption{\label{fig3}{\bf Linear stability analysis.} {\bf a-b} share the same $x$-axis and show the real and imaginary parts of the first two eigenvalues $\lambda_{1}$ (triangles) and $\lambda_{2}$ (circles) of the Jacobian $\mathcal{M}$ with fixed $J_x=5.9,~J_z=\gamma=1$. The nonzero conjugate pairs of the imaginary parts indicate oscillatory dynamics. This oscillation decays to a stationary spin-density-wave (SDW) phase when the real parts are negative, and stabilizes to a nonstationary oscillatory (OSC) phase when the real parts are positive. {\bf c-d} share the same $y$-axis and show the robustness of dynamics of $\langle\hat{\sigma}^{y}\rangle$ for different initial states $\ket{\psi_{n}(t=0)}=(\ket{\uparrow}+\sqrt{99}e^{i(r_n+c_n)\pi/k}\ket{\downarrow})/10$ with $k=3~\text{(black solid line)}, ~6~ \text{(blue short-dashed line)}, ~9~ \text{(red dotted line)}, ~12~ \text{(green dot-dashed line)}, ~15 ~\text{(magenta solid line)}$. The zoom-in plot of  {\bf d} and the Fourier spectrum of the selected dynamical regions ($\Gamma t\in[400,~1000]$) of  {\bf d}  are respectively shown in {\bf d1} and {\bf d2}. The dashed lines in {\bf a-b} highlight zero values.}  
\end{figure*}

Those OSC steady states can be further classified {into LC and chaos based on} the Lyapunov exponent $\lambda$ \cite{QLE2019chaos}. The LC steady states here are CTCs because they are characterized not only by stable periodic oscillations but also by the spontaneous breaking of continuous time-translation symmetry in many-body systems. For example, as shown in Fig. \ref{fig2}(b), we present a detailed phase diagram of OSC phases with fixed $J_{y}=1.1$, where the CTC and chaos correspond to $\lambda = 0$ and $\lambda > 0$, respectively (see details in the “Calculation of Lyapunov exponent” subsection in the "METHOD" section). In Figs. \ref{fig2}({c-g}), we demonstrate the different dynamical behaviors of FM, PM, SDW, CTC, and chaos phases by projecting the dynamic trajectories of magnetization
onto the $y$-$z$ plane of the Bloch sphere. The highlighted circular blocks correspond to the initial states $\ket{\psi_{n}(t=0)}=(\ket{\uparrow}+e^{in\pi/9}\ket{\downarrow})/\sqrt{2}$ for the spins with $n=1,~4,~7$ as labeled in Fig. \ref{fig1}(b). The dynamical behaviors of magnetization $\langle\hat{\sigma}^{\alpha=y,z}_{n=1,~4,~7}\rangle$
illustrate that the stationary  SDW phase corresponds to fixed points, while either LC attractors or chaotic behavior are observed in the nonstationary OSC phase.

The Heisenberg XXZ model is a loyal supporter of PM, where the system has isotropic interactions ($J_{x}=J_{y}$) in $x$-$y$ plane, which conserve magnetization, leaving no mechanism to counteract the dissipation-induced spin downwards. In the XYZ model, the anisotropic interaction ($J_{x}\neq J_{y}$) breaks the conservation of magnetization, {leading to} spin fluctuation with pairs of spins flipping upwards or downwards simultaneously. This spin fluctuation also allows each spin to precess around its effective magnetic field ${\vec{B}}^{\text{eff}}_{n}=\sum_{\langle mn\rangle}(J_{x}\langle\hat{\sigma}^{x}_{m}\rangle, J_{y}\langle \hat{\sigma}^{y}_{m}\rangle, J_{z}\langle \hat{\sigma}^{z}_{m}\rangle)$, which originates from interactions with surrounding spins as shown in Fig. \ref{fig1}(c). The competition between the spin fluctuation or precession and the dissipation-induced spin downwards complicates the scenario, leading to the emergence of other phases when the dissipation no longer dominates.  The balanced competition results in stationary phases such as FM and SDW. This balanced competition can also turn into cooperation under suitable parameters, giving rise to nonstationary OSC phases that break energy conservation.  Those steady states can only be achieved with sufficiently strong anisotropic interactions. Therefore, assuming a sufficiently large cluster size is essential for accessing a broader range of possible steady states. Figure \ref{fig1}(d) illustrates the case that neighboring spins tend to align antiparallel, i. e., $J_{\alpha=x,y,z}>0$ (other cases can be analyzed similarly), to illustrate the microscopic cycle of this dynamical cooperation. The system gains energy when the dissipation flips down the $n$th spin.  This process also would be recognized as implicit incoherent drive \cite{minganti2020correspondence} in the sense that it helps system to gain the energy. Following is an energy-conserving process that spin fluctuations cause two neighboring spins to be in a coherent superposition of up or down simultaneously. Finally, when dissipation flips the $(n+1)$th spin downward to {align} antiparallel with neighboring spins, the system releases energy.
  
The energy gain and loss of Floquet time crystals primarily arises from coherent driving. In driven-dissipative systems, the driving and dissipation can respectively serve as mechanisms for energy gain and loss \cite{minganti2020correspondence,Physics2021,PRL2018Gong,Zhu_2019,PRR2020DissFTC,RieraCampeny2020timecrystallinityin,PRB2021DissFTC,PRL2022seeding,PRL2023CTC,PRA2019dissTC,Tucker_2018,PRB2019DissCTC,PRA2020DissCTC,PRB2021dbtc,PRL2018BTC,PRL2021dissDTC,PRL2021Kongkhambut,Science2022CTC,Science2019Nishant,PRL2019Zupancic,Dreon2022Nature,taheri2022all,PRL2019CTCBEC,PRB2019CTC}. Without driving, if the dissipation solely reduces the system's energy, such as photon leakage from a cavity leading to irreversible energy loss, the system would decay into a trivial steady state \cite{PRL2021dissDTC,PRL2021Kongkhambut,Science2022CTC,Science2019Nishant,PRL2019Zupancic,Dreon2022Nature}. 
In our scenario, these microscopic cycles sustain the nonconservative OSC behavior, where dissipation plays the dual role of energy gain and loss in cooperation with anisotropic-interaction-induced spin fluctuations. Therefore, our work introduces,  to the best of our knowledge, an entirely new mechanism for the emergence of CTC.

\subsection*{Linear stability analysis}
To further confirm the time-dependent nonequilibrium phase, we introduce small local perturbations $\delta\hat{\rho}_{C}$ to the fixed-point solution $\hat{\rho}_{C}^{(0)}$ of {Eq. (\ref{BEs})}, which is numerically obtained by setting the left {hand} side of {Eq. (\ref{BEs})} to zero. The characteristic {equation of motion} for $\delta\hat{\rho}_{C}$ is derived by linearizing the system and denoted as $\partial_{t}\delta\hat{\rho}_{C}=\mathcal{M}[\delta\hat{\rho}_{C}]$, where the superoperator $\mathcal{M}$ is referred {to} as Jacobian. Therefore, the dynamical behavior of local perturbations can be expressed as $\delta\hat{\rho}_{C}=\sum_{j}c_{j}e^{\lambda_{j}t}\delta\hat{\rho}_{C,j}$, where $c_{j}$ and $\lambda_{j}$ are the superposition coefficients and eigenvalues corresponding to the $j$-th eigenmode $\delta\hat{\rho}_{C,j}$ of the Jacobian. Here, eigenvalues are {ordered in descending real parts}, i.e., $\text{Re}[\lambda_{1}]\geq\text{Re}[\lambda_{2}]\geq\cdots$. Therefore, when all eigenvalues have negative real parts, $\delta\hat{\rho}_{C}$ will decay to zero over the system's relaxation time, indicating that the fixed-point solution $\hat{\rho}_{C}^{(0)}$ represents the steady state. The appearance of eigenvalues with positive real parts results in the exponential growth of the local perturbations $\delta\hat{\rho}_{C}$, revealing that the fixed-point solution $\hat{\rho}_{C}^{(0)}$ is a metastable state. In additon, the system will evolve into a time-dependent steady state if the imaginary parts are nonzero.

Figure \ref{fig3}(a-b) shows the real and imaginary parts of the first two eigenvalues, $\lambda_{1}$ (triangles) and $\lambda_{2}$ (circles), of the Jacobian $\mathcal{M}$. The phase boundaries of the SDW-OSC phases, $J_y=1.098$ and $J_y=1.366$, where the real parts of the eigenvalues switch signs, are consistent with those shown in Fig. \ref{fig2}(a). The imaginary parts consistently appear as nonzero conjugate pairs, suggesting that by selecting an appropriate initial state, the system will: 1) {evolve into} the time-independent steady state with oscillatory decay behavior in the SDW phase; 2) potentially pass through a metastable state before transitioning into a {LC} state in the OSC phase. Those behaviors are illustrated in Figs. \ref{fig3}(c-d) by the average magnetization of one cluster, $\langle\hat{\sigma}^{\alpha=x,y,z}\rangle=\sum_{n\in C}\langle\hat{\sigma}_{n}^{\alpha=x,y,z}\rangle/N_C$ with initial state $\ket{\psi_{n}(t=0)}=(\ket{\uparrow}+\sqrt{99}e^{i(r_n+c_n)\pi/k}\ket{\downarrow})/10$, where $r_n$ and $c_{n}$ denote the row and column of the $n$th spin.  In Figs. \ref{fig3}(c-d), the lines represent results obtained from different initial states with $k=3,~6,~9,~12,~15$. These results demonstrate the robustness of the steady states against variations in initial conditions. Specifically, for the CTC phase, different initial states only induce a phase shift in the oscillations, as shown in Fig. \ref{fig3}(d1), while the amplitude and frequency spectrum, as depicted in Fig. \ref{fig3}(d2), remain unchanged.

\subsection*{Rigidity of continuous time crystal}
\begin{figure}
\centering
\includegraphics[width=8.cm]{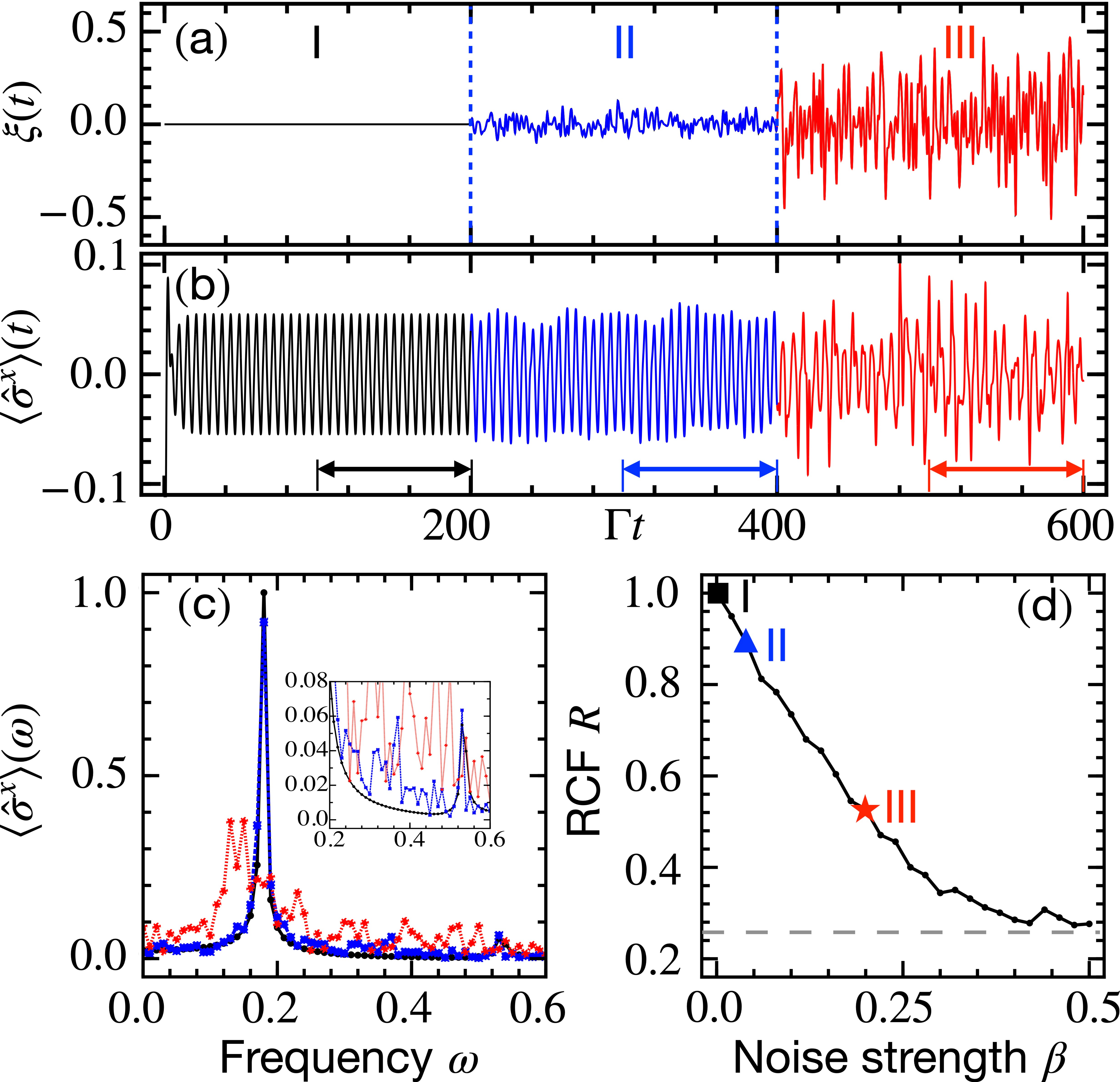}
\caption{\label{fig4} {\bf Rigidity of continuous time crystals (CTCs) under Gaussian white noise.} \textbf{a-b} share the $x$-axis and show the noisy interaction and spin dynamics with three stages: The noiseless CTC (stage I) is perturbed by weak noise (stage II) and strong noise (stage III) at \( \Gamma t = 200 \) and \( 400 \), respectively. The noise strengths $\beta_{\text{I}}=0,~\beta_{\text{II}}=0.04,~\beta_{\text{III}}=0.2$ are respectively indicated in {\bf d} with square, triangle and star symbols. {\bf c} shows the Fourier spectrum of the selected arrowed regions in {\bf b}, with the zoom-in inset. {\bf d} shows that the relative crystalline fraction (RCF)  $R$ decreases as the noise strength $\beta$ increases and stabilizes at the dashed gray line, where the system may still retain some information about the CTC. Here the initial state is the same as the one used in Fig. \ref{fig2}(c-e), and the undisturbed interaction strengths are $J_{x}=7,~J_{y}=1.5,~J_{z}=1$.}
\end{figure}

 {The rigidity of a CTC requires its periodic OSC behavior to exhibit strong robustness against fluctuations in system parameters. Here, we consider noisy interactions given by $\tilde{J}_{\alpha}(t)=J_{\alpha}+\xi_{\beta}(t)$, where $\xi_{\beta}(t)$ represents isotropic {Gaussian} white noise, and its standard deviation $\beta$, acts as intensity. In Fig. \ref{fig4}(a), we design three stages of interactions, {with the corresponding dynamical behavior displayed in Fig. \ref{fig4}(b) for $\langle \hat{\sigma}^{x} \rangle$, and its Fourier spectrum, $\langle \hat{\sigma}^{x} \rangle (\omega)$, for the selected regions, shown in Fig. \ref{fig4}(c).} Without noise (stage I), the system relaxes to a LC-type oscillation, {whose Fourier spectrum exhibits a} sharp dominant peak at $\omega=\omega_p=0.18$ and a small peak at $\omega\approx3\omega_p$, indicating the emergence of a prototypical CTC. {We then perturb the CTC with the weak noise of} strength $\beta=0.04$ at stage II. This perturbation has nearly no effect on the main peak, only introducing weak higher-frequency components in the Fourier spectrum as shown in the inset of Fig. \ref{fig4}(c), indicating that the CTC is still thriving. Finally, we quench the system with much stronger white noise of strength $\beta=0.2$ at stage III. The regular periodic oscillations are significantly altered, with the $\omega_p$ component no longer being the dominant peak, and numerous other frequency components appearing. Consequently, the CTC becomes softened under this strong noise.

Furthermore, we define the relative crystalline fraction (RCF), $R_{\eta}=\Omega_{\eta}/\Omega_{\text{I}}$, to quantify the rigidity of the CTC. The crystalline fraction for stage $\eta=\text{I,II,III}$ is defined as  $\Omega_{\eta}=\sum_{\omega\in[\omega_{p}-\delta\omega,\omega_{p}+\delta\omega]}\langle\hat{\sigma}_{\eta}^{x}\rangle(\omega)/\sum_{\omega\in[0,\Delta\omega]}\langle\hat{\sigma}_{\text{I}}^{x}\rangle(\omega)$, with the principal frequency $\omega_{p}$ {and its frequency width $\delta\omega$ in the noiseless} case. {Here $\Delta\omega$ denotes the frequency window used to calculate  the crystalline fraction.} We set $\delta\omega=0.08$ and $\Delta\omega=0.6$ for our calculations. The RCFs for the three stages are $R_{\text{I}}=1$, $R_{\text{II}}=0.89$, and $R_{\text{III}}=0.53$ as shown in Fig. \ref{fig4}(d), where illustrates how the CTC melts as {the }noise strength increases. The CTC's RCF decreases {as the noise increases} and stabilizes at the dashed gray line, indicating that, even in the presence of strong noise, the system's OSC behavior may still retain some information about the CTC. We have also verified its robustness under other noisy interactions (see Supplementary Note 3) and noisy dissipation.}

\subsection*{Experimental realization}
We have recently experimentally implemented independent loss and gain between $\ket{\downarrow}=\ket{F=0, m_{F}=0}$ and $\ket{\uparrow}=\ket{F=1, m_{F}=0}$ states in the $^{2}S_{1/2}$ manifold of $^{171}$Yb$^{+}$. This was achieved by optically pumping the ions to six auxiliary excited states and adiabatically eliminating the states exhibiting spontaneous emission \cite{PRR2023Zhang}.  
The XYZ interactions have been realized in experiments by coupling the collective motion of ions to {their} internal states \cite{debnath2016demonstration,PRX2023ion,lu2019global,RMP2021ion,zhang2017observation1}. We find that the power-law decay characteristics of these interactions do not influence the existence of OSC behavior (see Supplementary Note 4). The XYZ interactions have also been proposed to be realizable through techniques such as two-photon resonance in systems {such as }Rydberg atoms, Rydberg-dressed atoms, and dipolar atoms or molecules \cite{Phase2013PRL_Lee}.

\section*{Conclusions} 
In summary, we have proposed a CTC mechanism resulting from the competition between dissipation-induced spin downwards and anisotropic-interaction-induced spin precession or spin fluctuation. In the past, the general understanding of CTCs primarily focused on the interplay of dissipation and driving-induced quantum coherence but gave less attention to the role of interaction. Therefore, this represents, to the best of our knowledge,  a completely new ergodic-breaking mechanism in which interactions play a crucial role. Since there is no external drive, the realized CTC is not subject to heating, allowing its lifetime can be as long as the system's.

This work paves the way for future explorations of CTCs that are immune to heating in other systems, including  Hubbard model \cite{PRA2016Wilson,PRL2016Schir} and Rydberg model \cite{PRL2022Nill,PRL2023Kazemi,PRL2019Gambetta} using the methods introduced here and other numerical approaches \cite{CMF2016PRX, RMP2005scholl,PRL2015Cui,PRL2019Rota,PRL2015Fina} involving quantum fluctuations and  correlations. The experimentally accessible CTC provides further motivation to use dissipations as resources for exploring nonequilibrium quantum dynamics, such as quantum synchronization. The existence of CTCs and chaotic phases opens a pathway to investigate reversible-to-irreversible transitions \cite{reichhardt2022perspective} in dissipative quantum systems.

\section*{Method}
\subsection*{Calculation of Lyapunov exponent }
\begin{figure}[H]
\centering
\includegraphics[width=8.cm]{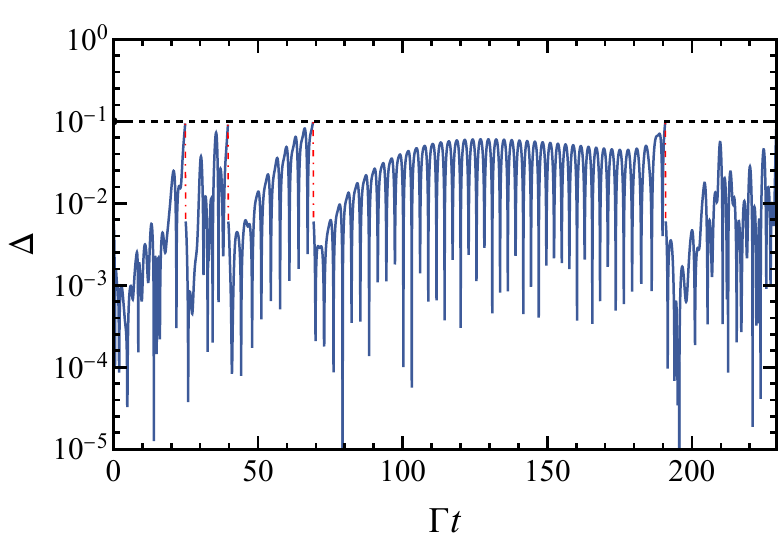}
\caption{{\bf Dynamics of distance.} Time evolution of the distance $\Delta(t)$ between the fiducial trajectory and auxiliary trajectory for chaotic phase. The black dashed horizontal line denotes the given threshold $\Delta_{\text{max}}=0.1$ for resetting the auxiliary trajectory (red dot-dashed vertical lines). Here $J_x=13, ~J_y=1.1, ~J_z=\gamma=1.0$.}\label{fig5}
\end{figure}
We employ the largest Lyapunov exponent (LE), a metric introduced in \cite{QLE2019chaos}, to make discrimination between {LC} and chaos behavior. In analogy to the classical definition, we use two quantum trajectories, fiducial trajectory and auxiliary trajectory, to simulate the evolution of the system.  
Here the auxiliary trajectory is initialized as a normalized state but with a perturbation on the normalized initial fiducial state $\psi_f^{\text{ini.}}$, namely $\psi_a^{\text{ini.}}=\psi_f^{\text{ini.}}+\varepsilon\psi_{\text{ran.}}$ with the random perturbative state $\psi_{\text{ran.}}$ and $\varepsilon\ll1$. The definition of the largest LE is based on the distance of these two trajectories and the distance is defined as difference between observables of these two trajectories. Here we choose the magnetization along $y$ direction $\overline{\sigma}^y(t)=\frac{1}{N_C}\sum_{n\in C}\langle\psi_n(t)|\hat{\sigma}^y(t)|\psi_n(t)\rangle$ as the observable. 
The initial distance $\Delta(t=0)=|\overline{\sigma}^y_f(t=0)-\overline{\sigma}^y_a(t=0)|$ plays the reference during the time evolution. When the time-dependent distance $\Delta(t)=|\overline{\sigma}^y_f(t)-\overline{\sigma}^y_a(t)|$ exceeds a given threshold $\Delta_{\text{max}}$ at a time $t_{k}$, $\Delta(t_k)$ is reset to the initial distance $\Delta(t=0)$ through renormalizing the auxiliary trajectory close to the fiducial trajectory.  Finally, the largest LE is given by
\begin{equation}
\lambda=\lim\limits_{t\rightarrow\infty}\frac{1}{t}\sum_k {\rm ln}\,d_k,
\end{equation}
where $k$ indexes each time point that the threshold $\Delta_{\text{max}}$ is touched and $d_k=\Delta(t_k)/\Delta_0$. The fact that auxiliary trajectory respectively tends to {be} attracted to the fiducial state in the {LC} case and tends to away from the fiducial state in the chaos behavior,  results in vanished LE $\lambda=0$ and {nonzero} LE  $\lambda\neq0$ for {LC and chaotic} behavior respectively.

As shown in Fig. \ref{fig5}, here we consider the parameters as $J_x=13, ~J_y=1.1, ~J_z=\gamma=1.0$, corresponding to the chaotic phase in the Fig. 2(b) of the main text. It can be observed that the dynamical distance $\Delta$ frequently touches the set threshold distance $\Delta_{\text{max}}$ during the evolution, indicating the {nonzero} LE $\lambda=0.1$ for the corresponding chaotic scenario. \\

\noindent
{\bf Data availability.}\\
The data that support the plots within this paper and other findings of this study are
available from the authors upon reasonable request.\\

\noindent
{\bf Code availability.}\\
The computer codes that support the calculations within this paper are available upon reasonable request via email to S. Yang. or J. Jie.\\

\noindent  
{\bf Acknowledgments}\\
We thank Ran Qi, Qingze Guan, Zhiyuan Sun, Jiansong Pan, Jin Zhang and Weidong Li for valuable discussions. We especially acknowledge Augusto Smerzi for his useful comments and advice on our manuscript. This work was supported by the National Natural Science Foundation of China (Grant No. 12104210, 12088101 and U2330401), the Natural Science Foundation of Top Talent of SZTU (GDRC202202, GDRC202312), and the Guangdong Provincial Quantum Science Strategic Initiative (No. GDZX2305006).\\

\noindent 
{\bf Author contributions}\\
J. Jie initiated this project. S. Yang carried out the theoretical derivations and numerical simulations. Z. Wang provided support for the numerical techniques. J. Jie and L. Fu took on advisory roles. All authors contributed to the  discussion and writing of the manuscript.\\

\noindent 
{\bf Correspondence} and requests for materials should be addressed to Libin Fu or Jianwen Jie.\\

\noindent 
{\bf Competing interests}\\
The authors declare no competing interests.

\global\long\def\id{\mathbbm{1}}
\global\long\def\ui{\mathbbm{i}}
\global\long\def\ud{\mathrm{d}}

\setcounter{equation}{0} \setcounter{figure}{0}
\setcounter{table}{0} 
\renewcommand{\theparagraph}{\bf}
\renewcommand{\thefigure}{S\arabic{figure}}
\renewcommand{\theequation}{S\arabic{equation}}
\renewcommand{\figurename}{Supplementary Figure}
\renewcommand{\eqref}[1]{Supplementary Equation~\ref{#1}}

\onecolumngrid
\flushbottom
\newpage
\section*{Supplementary Material for "Emergent Continuous Time Crystal in Dissipative Quantum Spin System without Driving"}

In this Supplemental material, we provide more details on (1) Two-body quantum correlations in finite-sized systems; (2) Nonequilibrium phase diagram of the systems cluster sizes $2\times2$ and $4\times4$;  (3) The robustness of CTC under $1/f$ noise; (4) The oscillatory behavior under power-law decay interaction;.

\section*{Supplementary Note 1: Two body quantum correlations in finite-sized systems}
In our mean-field calculations, all spins are assumed to form the entire system in a product state, so correlations between spins are not considered. Typically, directly considering quantum correlations in the thermodynamic limit is challenging. To explore the correlations in our system, we start by examining the two-body correlations in a finite-size system and study how these correlations change as the system size increases.

Specifically, we consider the parameters in the OSC region of Fig. 2a in the main text, denoted as $J_{x}=5.9,~J_{y}=1.2~, J_{z}=\gamma=1.0$. We examine the two-body spin-spin correlations along {\color{black}the $z$-axis}, $\langle \hat{\sigma}_{j}^{z}\hat{\sigma}_{k}^{z}\rangle_{c}=\langle \hat{\sigma}_{j}^{z}\hat{\sigma}_{k}^{z}\rangle-\langle \hat{\sigma}_{j}^{z}\rangle\langle\hat{\sigma}_{k}^{z}\rangle$. When $\langle \hat{\sigma}_{j}^{z}\hat{\sigma}_{k}^{z}\rangle_{c}\neq0$, there are correlations between spins. For this finite-size system, we use a fully quantum method to solve the Lindblad equation \cite{JOHANSSON20131234} shown in Eq. (1) of the main text, and obtain the results shown in Supplementary Figure \ref{FigS1}.

As shown in Supplementary Figure \ref{FigS1}, we computed the steady states of two-dimensional  systems with sizes 3$\times$3, 3$\times$4, 3$\times$5, and 4$\times$4 under periodic boundary conditions. We then calculated the two-body correlation between the first spin (located at the first row and first column) and the second spin  (located at the first row and second column). We compare the results of the 3$\times$3, 3$\times$4, 3$\times$5, and 4$\times$4 systems. The numerical results indicate that as the size of the finite system increases, the two-body correlations {\color{black}significantly} weaken. This suggests that in the thermodynamic limit, the two-body correlations will not play {\color{black}a} key role. This prediction aligns with the approximations made in our mean-field method.

\begin{figure}[h]
\centering
\includegraphics[width=8.63cm]{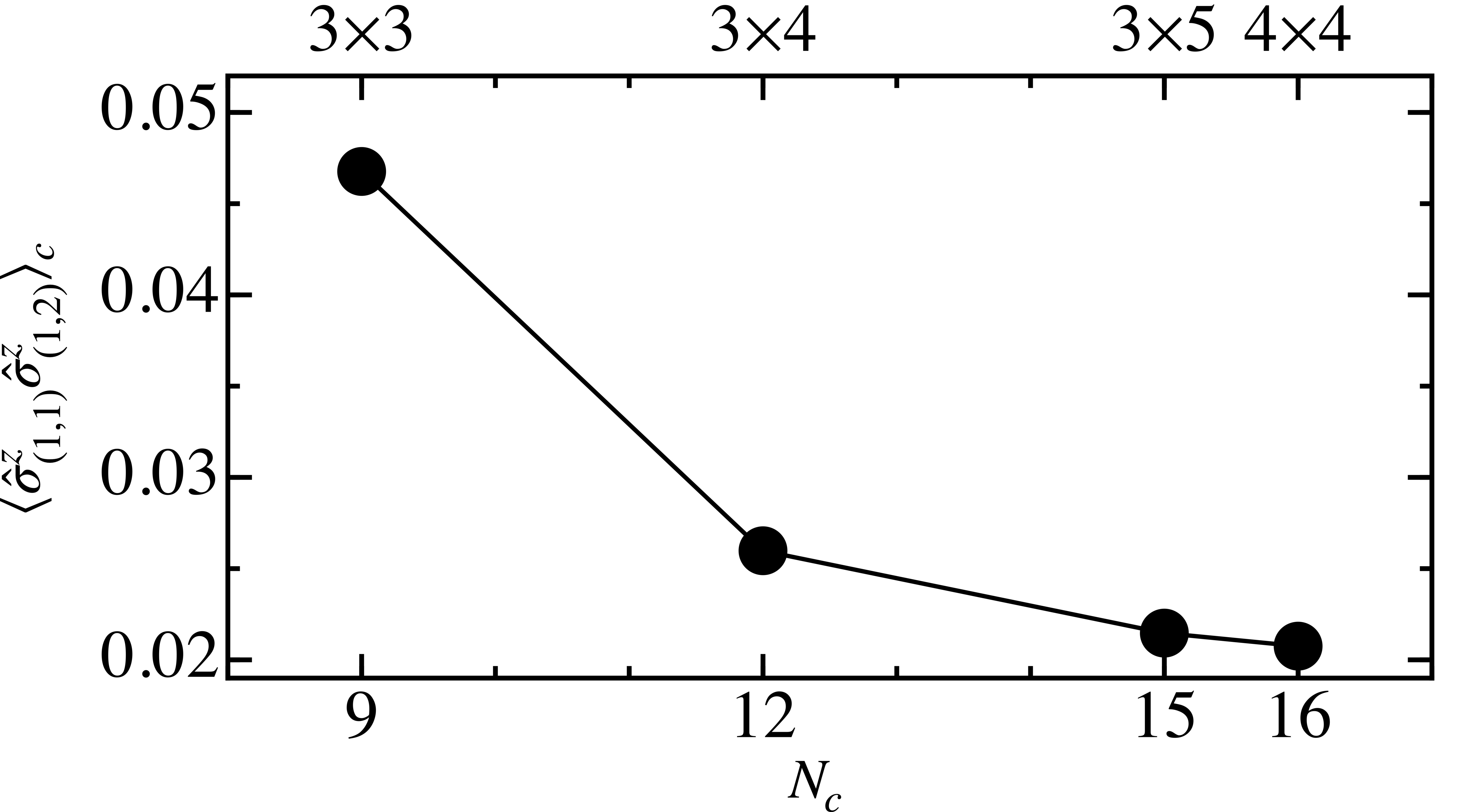}
\caption{{\bf Quantum correlation.}Two-body correlations are shown for different finite-size 2D systems (3$\times$3, 3$\times$4, 3$\times$5, and 4$\times$4). The notation $(r,c)$ denotes the position of a spin located at the $r$th row and $c$th column. The interaction parameters are set to $J_{x}=5.9,~J_{y}=1.2~, J_{z}=\gamma=1.0$.}
\label{FigS1}
\end{figure}

\section*{Supplementary Note 2: Nonequilibrium phase diagram of the systems  cluster sizes $2\times2$ and $4\times4$}
In our mean-field method, the system is assumed to be filled with identical clusters repeated throughout. Therefore, the symmetrical structure of a single cluster constrains the possible nonequilibrium steady states that can be achieved. For clusters of even length, such as $N_{c}=2$, an antiferromagnetic configuration can be constructed. However, for clusters of odd length, such as $N_{c}=3$, an antiferromagnetic configuration cannot be constructed. These analyses also apply to the spatial configuration of a staggering XY phase with an even period. The results shown in Supplementary Figure 2(a) in the main text are for the $3\times3$ cluster; hence, the antiferromagnetic and staggering XY phases as mentioned in Lee \textit{et al.} \cite{Phase2013PRL_Lee} cannot be constructed.

As illustrated in Supplementary Figure \ref{FigS2}(a-b), when we set the cluster size to $2\times2$, both the antiferromagnetic and staggering XY phases appear at $J_{z}=1$ [Supplementary Figure \ref{FigS2}(a)] and $J_{z}=0$ [Supplementary Figure \ref{FigS2}(b)], respectively. Supplementary Figure \ref{FigS2}(a) clearly shows that a $2\times2$ cluster cannot realize the SDW phase as described in Lee \textit{et al.} \cite{Phase2013PRL_Lee} which is consistent with the definition of SDW requiring a spin repetition period of greater than 2 in at least one direction. Supplementary Figure \ref{FigS2}(c) shows the results for a $4\times4$ cluster, where, in addition to the phases observed in Fig. 2(a) in the main text, the antiferromagnetic phase also appears.

\begin{figure*}
\centering
\includegraphics[width=13.63cm]{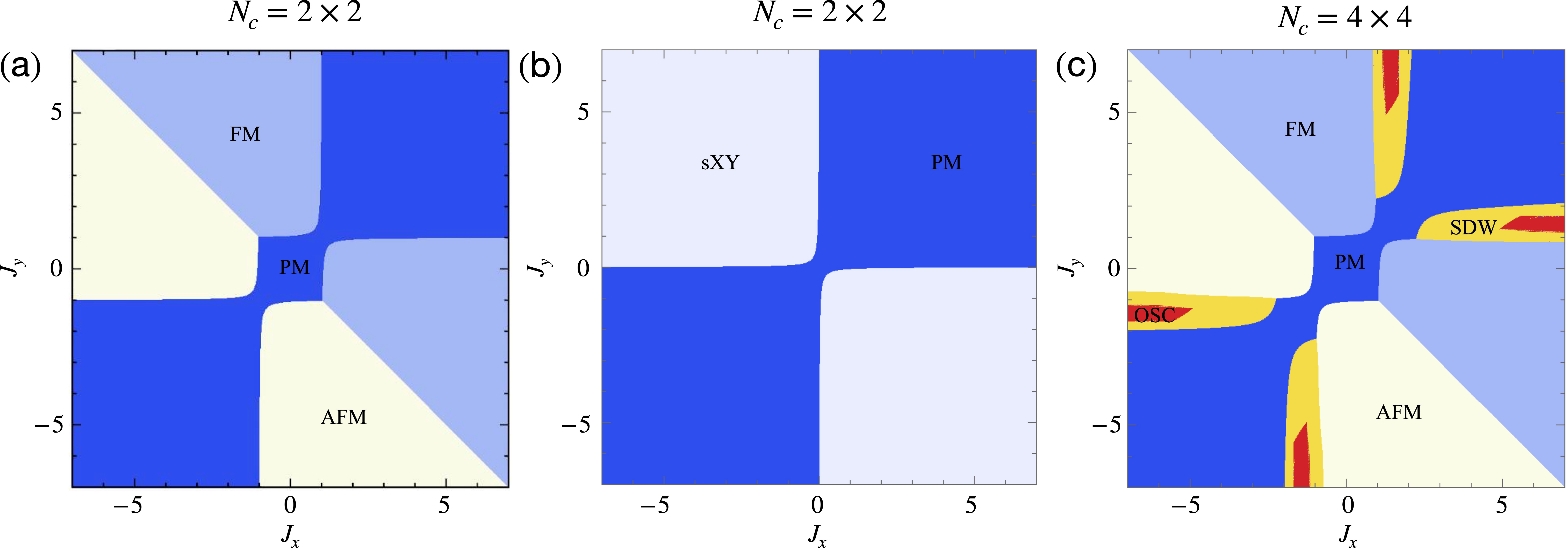}
\caption{{\bf Nonequilibrium phase diagram for other cluster sizes.} {\bf a-c} show the mean-field phase diagram for the cluster sizes $2\times2$ with $J_z=1$, $2\times2$ with $J_z=0$, and $4\times4$ with $J_{z}=\gamma=1$, respectively. Here includes five stationary phases: paramagnetic (PM), ferromagnetic (FM), antiferromagnetic (FM), staggering XY (sXY), and spin-density-wave (SDW), along with one additional nonstationary oscillatory (OSC) phase.}
\label{FigS2}
\end{figure*}

\section*{Supplementary Note 3: The robustness of CTC under $1/f$ noise}
\begin{figure}[H]
\centering
\includegraphics[width=8.6cm]{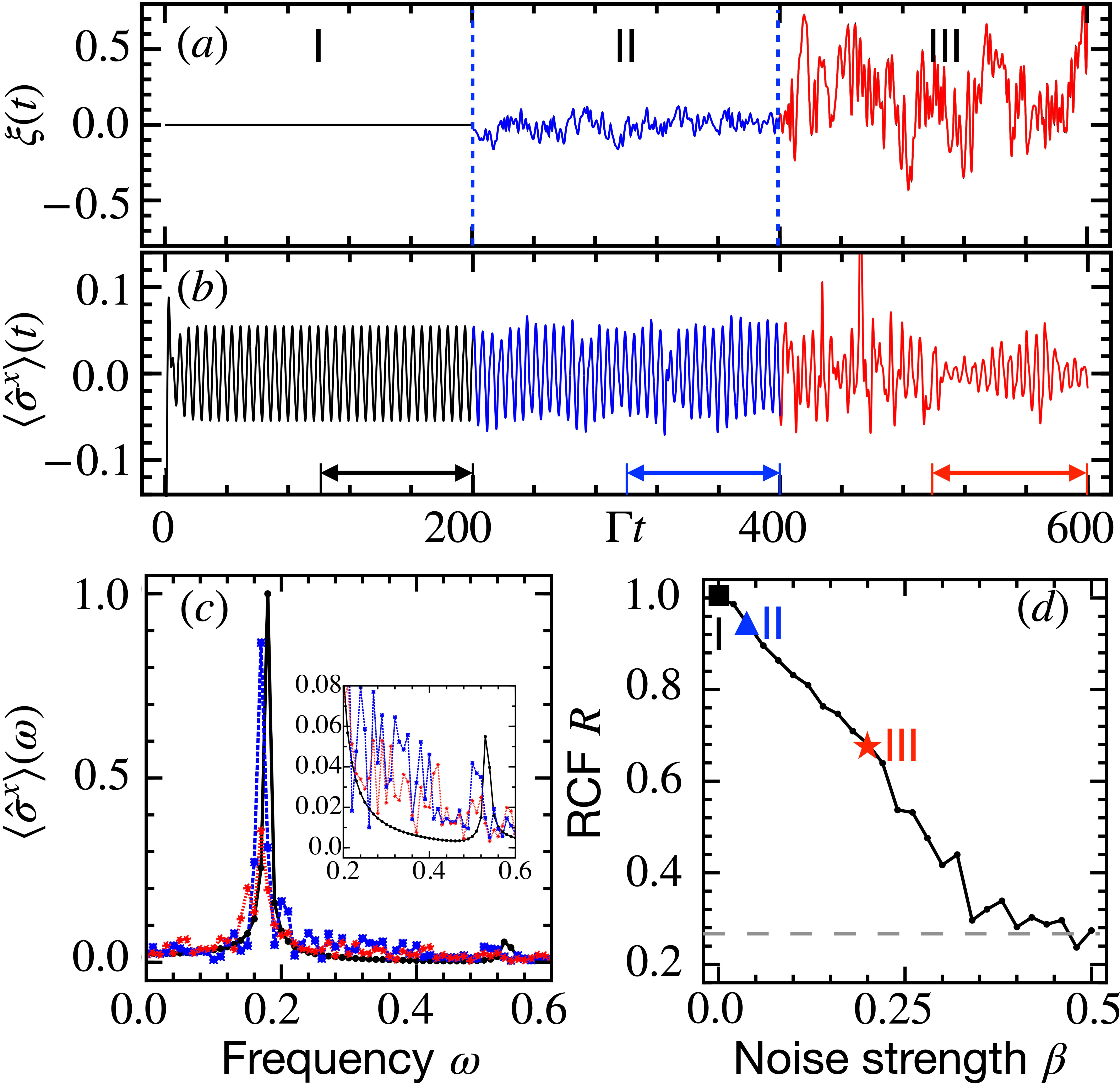}
\caption{{\bf Rigidity of continuous time crystals (CTCs) under $1/f$ noise.} \textbf{a-b} share the $x$-axis and show the noisy interaction and spin dynamics with three stages: The noiseless CTC (stage I) is perturbed by weak noise (stage II) and strong noise (stage III) at \( \Gamma t = 200 \) and \( 400 \), respectively. The noise strengths $\beta_{I}=0,~\beta_{II}=0.04,~\beta_{III}=0.2$ are respectively indicated in {\bf d} with square, triangle and star symbols. {\bf c} shows the Fourier spectrum of the selected arrowed regions in {\bf b}, with the zoom-in inset. {\bf d} shows that the relative crystalline fraction (RCF)  $R$ decreases as the noise strength $\beta$ increases and stabilizes at the dashed gray line, where the system may still retain some information about the CTC. Here the initial state is the same as the one used in Fig. 2(c-g) of the main text, and the undisturbed interaction strengths are $J_{x}=7,~J_{y}=1.5,~J_{z}=1$.}\label{FigS3}
\end{figure}
$1/f$ noise, often referred to as flicker noise or pink noise, represents a specific type of noise marked by a frequency spectrum that diminishes with increasing frequency. To clarify, the power spectral density of $1/f$ noise follows an inverse proportionality to the frequency, resulting in a higher concentration of power at lower frequencies. Here we consider the robustness of CTC under $1/f$ noise. The noisy interaction is given by $\tilde{J}_{\alpha}(t)=J_{\alpha}+\xi_{\beta}(t)$, where $\xi_{\beta}(t)$ represents $1/f$ noise and its standard deviation $\beta$ acts as intensity. 

Similar to the consideration of isotropic Gaussian-type white noise shown in Fig. 4 in the main text, here we show the results in Supplementary Figure \ref{FigS3}. Three stages of time-dependent interactions with $1/f$ noise are shown in Supplementary Figure \ref{FigS3}(a) and the corresponding dynamical behavior is displayed in Supplementary Figure \ref{FigS3}(b) with $\langle\hat{\sigma}^{x}\rangle$, whose Fourier spectra for the selected regions that away from relaxing processes are shown in Supplementary Figure \ref{FigS3}(c). Without noise (stage I), the system relaxes to a LC-type oscillation, the Fourier spectrum of which features a discernible sharp dominant peak at $\omega=\omega_p=0.18$ and a tiny peak at $\omega\approx3\omega_p$, indicating the {\color{black}emergence of} a prototypical CTC. Then we perturb the CTC by introducing the weak $1/f$ noise with strength $\beta=0.04$ at $\Gamma t=200$ (stage II), this perturbation slightly {\color{black}shifts} the dominant peak to the low frequency regime and introduces some weak higher-frequency components into the OSC behavior as illustrated in the inset of Supplementary Figure \ref{FigS3}(c), indicating that the CTC is still well {\color{black}alive}.
Finally, we quench the $1/f$ noise to a much stronger one with strength $\beta=0.2$ at $\Gamma t=400$ (stage III), which significantly alters the regular periodic oscillations. Interestingly, Although the amplitudes of the oscillating {\color{black}change} significantly, the prominent principal peak of frequency spectrum {\color{black}does not shift significantly}. Figure \ref{FigS3}(d) shows the relative crystalline fraction (RCF). The RCFs for three stages are respectively $R_{\text{I}}=1$, $R_{\text{II}}=0.94$, and $R_{\text{III}}=0.68$ as remarked in Supplementary Figure \ref{FigS3}(d), where illustrates how the CTC melts as noise strength increases. The CTC's RCF decreases with increasing noise and stabilizes at the dashed gray line, indicating that even in the presence of strong noise, the system's OSC behavior still may retain some information about the CTC.

\section*{Supplementary Note 4: The oscillatory behavior under power-law decay interaction}

In current trapped-ion platform \cite{debnath2016demonstration,PRX2023ion,lu2019global,RMP2021ion}, it is possible to achieve Heisenberg XYZ interactions. However, these interactions are power-law decay with distance, whereas the interactions discussed in our work are primarily nearest-neighbor XYZ interactions.  For the dissipation of spin-1/2 particles or qubits, well-established optical pumping methods can already achieve this. In our recent trapped-ion experimental work \cite{PRR2023Zhang}, we successfully realized a pair of independent dissipations, one is the decay process that spin jumps from $\ket{\uparrow}$ to $\ket{\downarrow}$ and another one is the gain process that spin jumps from $\ket{\downarrow}$ to $\ket{\uparrow}$, by optically pumping the ion from the targeted two lower energy levels to the six auxiliary excited energy levels. 

Although the trapped-ion interactions are power-law decay with distance, our calculations show that even this type of power law decaying interaction can induce oscillatory behavior as illustrated in Supplementary Figure \ref{FigS4}. To obtain those results, we start with setting the following Lindblad equation,
\begin{align}
\label{Rlindblad}
 \frac{d\hat{\rho}(t)}{dt}= -i\left[\hat{H}, \hat{\rho}(t)\right] +\frac{1}{2}\sum_n \mathcal{D}_{n}[\hat \rho(t)],
\end{align}
with the dissipation 
\begin{eqnarray}\label{Dj}
\mathcal{D}_{n}[\hat{\rho}] = \Gamma\left(\hat{\sigma}^{-}_{n}\hat{\rho} \hat{\sigma}^{+}_{n} - \{ \hat{\sigma}^{+}_{n}\hat{\sigma}^{-}_{n} , \hat{\rho} \}/2\right),
\end{eqnarray}
and  the Hamiltonian
\begin{eqnarray}\label{RHamiltonian}
	\hat{H}=\frac{1}{2d}\sum_{m,n}\sum_{\alpha=x,y,z}\frac{J_\alpha}{|r_m-r_n|^{\beta}}\hat{\sigma}^\alpha_m\hat{\sigma}^\alpha_n,
\end{eqnarray}
where $r_{m(n)}$ is the position of $m(n)$th spin and the exponent $\beta$ determines the range of the interaction can be set to $0<\beta<3$ by adjusting the laser detunings \cite{debnath2016demonstration,PRX2023ion,lu2019global,RMP2021ion}. Applying the mean-field method used in our work, we obtain the following set of $3N_c$ coupled nonlinear Bloch equations
\begin{align}\label{RBEs}
\frac{d\langle\hat{\sigma}^x_n\rangle}{dt}&=-\frac{\Gamma}{2}\langle\hat{\sigma}^x_n\rangle+\frac{1}{d}\sum_{m\neq n}\frac{J_y\langle\hat{\sigma}^z_n\rangle\langle\hat{\sigma}^y_m\rangle-J_z\langle\hat{\sigma}^y_n\rangle\langle\hat{\sigma}^z_m\rangle}{|r_{m}-r_{n}|^{\beta}},\\
\frac{d\langle\hat{\sigma}^y_n\rangle}{dt}&=-\frac{\Gamma}{2}\langle\hat{\sigma}^y_n\rangle+\frac{1}{d}\sum_{m\neq n}\frac{J_z\langle\hat{\sigma}^x_n\rangle\langle\hat{\sigma}^z_m\rangle-J_x\langle\hat{\sigma}^z_n\rangle\langle\hat{\sigma}^x_m\rangle}{|r_{m}-r_{n}|^{\beta}},  \\
\frac{d\langle\hat{\sigma}^z_n\rangle}{dt}&=-\Gamma(\langle\hat{\sigma}^z_n\rangle+1)+\frac{1}{d}\sum_{m\neq n}\frac{J_x\langle\hat{\sigma}^y_n\rangle\langle\hat{\sigma}^x_m\rangle-J_y\langle\hat{\sigma}^x_n\rangle\langle\hat{\sigma}^y_m\rangle}{|r_{m}-r_{n}|^{\beta}}.
\end{align}
Thus, the results in Supplementary Figure \ref{FigS4} can be obtained by numerically solving the above coupled nonlinear Bloch equations. In Supplementary Figure \ref{FigS4}, we set $\beta=1,J_{y}=9,~J_{z}=\gamma=1$ and use the initial state from Fig. 3(b-c) in the main text. We find that OSC oscillations occur when the interaction is anisotropic (Supplementary Figure \ref{FigS4}(a), $J_{x}=0\neq J_{y}$). However, for isotropic interactions (Supplementary Figure \ref{FigS4}(b), $J_{x}=J_{y}=9$), the system evolves to a non-OSC steady state. Therefore, we believe that the current trapped-ion platform already has the experimental conditions necessary to observe nonequilibrium oscillatory behavior.

\begin{figure*}
\centering
\includegraphics[width=12.63cm]{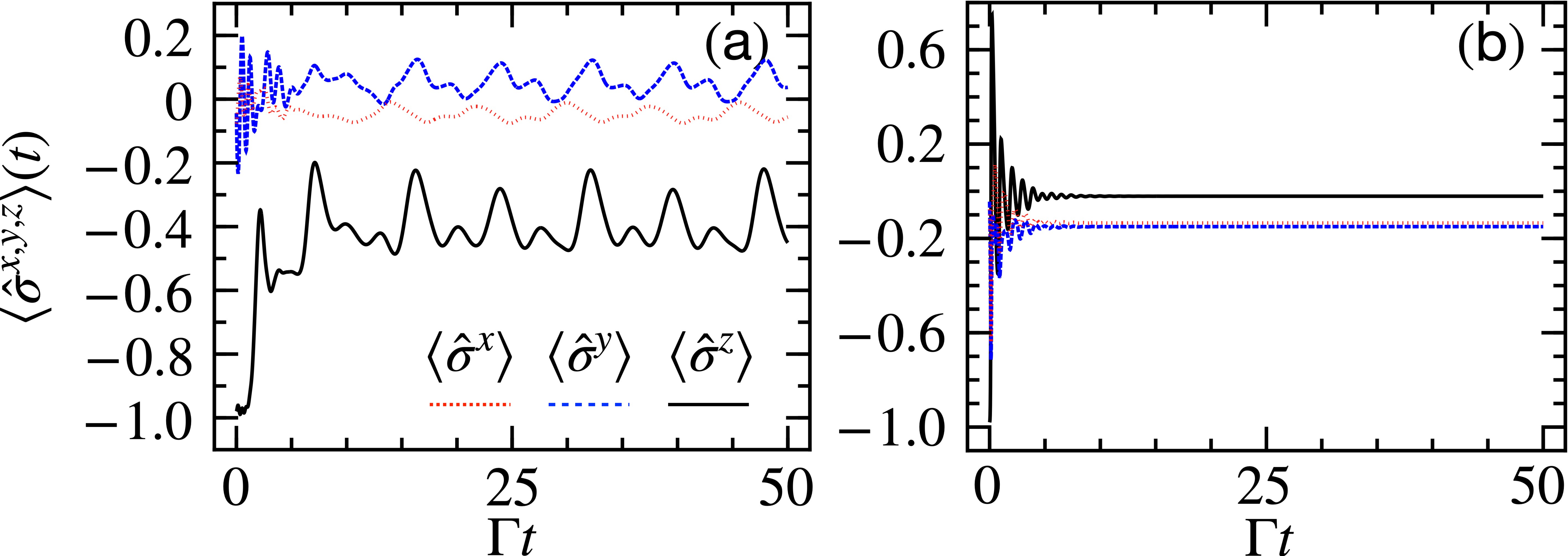}
\caption{{\bf Power-law decay XYZ interaction.} {\bf a} and {\bf b} show the spin dyncamics of a OSC-type steady state ($J_{x}=0$) and a non-OSC-type steady state ($J_{x}=9$) for the dissipated Heisenberg spin systems with power-law decay XYZ interaction, respectively.  The parameters are set to $J_{y}=9,~J_{z}=\gamma=1,~\beta=1$.  The initial state is the same as the one used in Fig. 3(b-c) of the main text. Both panels share the same y-axis label. }
\label{FigS4}
\end{figure*}

\noindent {\bf Supplementary References}

\end{document}